# Spatio-temporal isolator in lithium niobate on insulator

Haijin Huang[1], Armandas Balčytis[1], Aditya Dubey[1], Andreas Boes[1,2,3], Thach G. Nguyen[1], Guanghui Ren[1], Mengxi Tan[1], and Arnan Mitchell[1]

[1] *Integrated Photonics and Applications Centre, School of Engineering, RMIT University, Melbourne, VIC 3001, Australia*

[2]*School of Electrical and Electronic Engineering, The University of Adelaide, Adelaide, SA 5005, Australia*

[3]*Institute for Photonics and Advanced Sensing, The University of Adelaide, Adelaide, SA 5005, Australia*
arnan.mitchell@rmit.edu.au

**Abstract:** In this contribution, we simulate, design, and experimentally demonstrate an integrated optical isolator based on spatiotemporal modulation in the thin-film lithium niobate on insulator waveguide platform. We used two cascaded travelling wave phase modulators for spatiotemporal modulation and a ring resonator as a wavelength filter to suppress the sidebands of the reverse propagating light. This enabled us to achieve an isolation of 27 dB. The demonstrated suppression of the reverse propagating light makes such isolators suitable for the integration with III-V laser diodes and Erbium doped gain sections in the thin-film lithium niobate on insulator waveguide platform.

# 1. Introduction

Thin-film lithium niobate on insulator (LNOI) is a photonic integrated circuit platform that has enabled breakthrough demonstrations such as optical frequency comb generation [1-3] and second harmonic generation [4-6] in recent years. These achievements are enabled by lithium niobate's standout material properties, which include a strong optical nonlinearity, its piezoelectricity, and the ability to achieve phase modulation with minimal amplitude modulation via the electro-optic effect [7]. Recently, light sources such as amplifiers [8] and lasers [9] have also been integrated in LNOI by using Erbium doping and the integration of III-V laser diodes [10]. However, such integrated light sources are sensitive to back reflections, which can cause detrimental instabilities [11]. Reduction of such undesired feedback into the light sources to low enough levels so that they do not interfere with their operation can be achieved through the use of non-reciprocal devices such as optical isolators and circulators. Nonreciprocal optical devices have been demonstrated in several other photonic integrated circuit platforms by using three main methods: magnetic biasing [12], optical nonlinearity [13], and spatiotemporal modulation [14]. Magnetic biasing is attractive due to its broadband nature, but requires the integration of magneto-optical materials, which is a challenging fabrication process and induced high optical losses [15]. Optical nonlinear non-reciprocal devices are possess an advantage in that they can be achieved monolithically in LNOI. However, their operation is dependent on the power of input light sources [13], which is not always desired. Isolators that use spatiotemporal modulation operate independent of the optical power from the light source and appear to be an attractive way to achieve isolators in LNOI monolithically, by leveraging the excellent electro-optical modulation characteristics of LNOI [16].

In this contribution, we simulate, design and experimentally demonstrate integrated isolators in the LNOI waveguide platform. The nonreciprocal operation is achieved by using spatiotemporal modulation of two cascaded travelling wave phase modulators. The microwave signal that was applied to the modulators was selected so that most of the optical power is transferred to the sidebands when the device is operated in the reverse direction. The sidebands are then suppressed by an integrated add-drop ring resonator, enabling us to demonstrate an optical isolation of 27 dB.

## 2. Device design

Fig. 1 (a) shows the schematic overview of our proposed integrated isolator in LNOI, which consists of two identical travelling wave phase modulators connected in series and an add-drop ring resonator for spectral filtering. Like the tandem phase modulator-based optical isolator in the silicon on insulator waveguide platform [17], two phase modulators are modulated with two microwave signals having the same amplitude A but with a phase difference φ. In the forward direction, the output optical field after passing through the two modulators can be written as:

$$
\begin{aligned}
E_f &= E_o e^{i\omega_c t} e^{iR_f \cos(2\pi f_s t + \phi)} e^{iR_f \cos(2\pi f_s(t + \Delta T))} \\
&= E_o e^{i\omega_c t} \; e^{i2R_f \cos(2\pi f_s t + (2\pi f_s \Delta T + \phi)/2)\cos((2\pi f_s \Delta T - \phi)/2)}
\end{aligned}
\tag{1}
$$

where $E_o$ and $\omega_c$ are the optical carrier's amplitude and angular frequency, $\Delta T$ is the time delay between the two modulators, $R_f = A/V_{\pi,f}$ is the modulation index, and $V_{\pi,f}$ is the modulator switching voltage for the forward direction. When the phase difference φ and the time delay $\Delta T$ satisfy the following condition:

$$
2\pi f_s \Delta T - \phi = \pi + k\pi
\tag{2}
$$

where k is an integer, the output field is identical to the input field for the forward direction, which means that the optical field is unchanged, regardless of the modulation signal power. However, when the light passes through the modulation area in reverse, the output optical field will be:

$$
E_b = E_o e^{i\omega_c t} e^{iR_b \cos(2\pi f_s t)} e^{iR_b \cos 2\pi f_s(t + \Delta T + \phi)}
$$

$$
= E_o e^{i\omega_c t} e^{i2R_b \cos\left(2\pi f_s t + \frac{2\pi f_s \Delta T + \phi}{2}\right)\cos(\phi + 2\pi f_s \Delta T)/2)}
$$

$$
= E_o e^{i\omega_c t} \sum_{n=-\infty}^{n=+\infty} i^n J_n\left(\frac{2R_b \cos(\phi + 2\pi f_s \Delta T)}{2}\right) e^{in\left(2\pi f_s t + \frac{2\pi f_s \Delta T + \phi}{2}\right)}
\tag{3}
$$

where $R_b$ is the modulation index when the light is propagation in the reverse direction, $J_n(x)$ represents the $n$th order Bessel function of the first kind. When operating in the reverse direction, the power of the optical carrier is transferred to the sidebands, and the remaining power in the carrier is proportional to $J_n(2R_b \cos(\phi + 2\pi f_s \Delta T)/2)$. The sidebands can be filtered out by the add-drop ring resonator, leaving only the carrier passing through. To ensure strong suppression of the carrier and transfer of most of the optical field to the sidebands, we need to operate the modulators for the reverse-propagating light such that modulation index of the two modulators:

$$2R_b \cos\left(\frac{\phi + 2\pi f_s \Delta T}{2}\right) \approx 0.77\pi \tag{4}$$

which results in the carrier $J_0(2R_b \cos(\phi + 2\pi f_s \Delta T)/2) \approx 0$. Combining (2) and (4), the conditions for isolation operation are [18]:

$$2\pi f_s \Delta T = \pi + k\pi + \phi \quad \text{and} \quad R_b = \frac{0.77\pi}{2\sin\left(\frac{k\pi}{2} - 2\pi f_s \Delta T\right)} \tag{5}$$

In this work, we select $\phi = \pi/2$ and k=2, resulting in $\Delta T = 7/4f_s$ to minimize the requirement on the RF driving power [18].

The photonic waveguide platform that is used in this work to demonstrate the isolator is the silicon nitride loaded LNOI [19, 20]. A cross-section of the phase modulator is illustrated in Fig. 1(b). For the operation of the modulators, we chose a modulation frequency $f_s = 22.5$ GHz and of delay time was $\Delta T = 7/4f_s$, which, for a group index of 2.23, corresponds to a waveguide length of 10.4 mm. When the modulators are driven with a $\pi/2$ phase difference, the simulated spectra for the forward and reverse propagating light can be seen in Fig. 1(c) and (d), respectively. One can see that for the forward direction all of the optical power remains at the carrier wavelength, whereas in reverse direction most of the optical power is transferred to the sidebands, while the power of the carrier wavelength (1550 nm) is strongly reduced. To suppress the sidebands in the reverse direction we designed an add-drop ring resonator with a radius of 200 µm, resulting in a free spectral range of 55 GHz. For estimating the sharpness of the resonances of the ring resonator, we assumed a waveguide loss of 1 dB/cm. The expected add-drop filter transmission spectrum can be seen as the purple line in Figs. 1(c) and (d), which should be able to suppress most of the power in the sidebands.

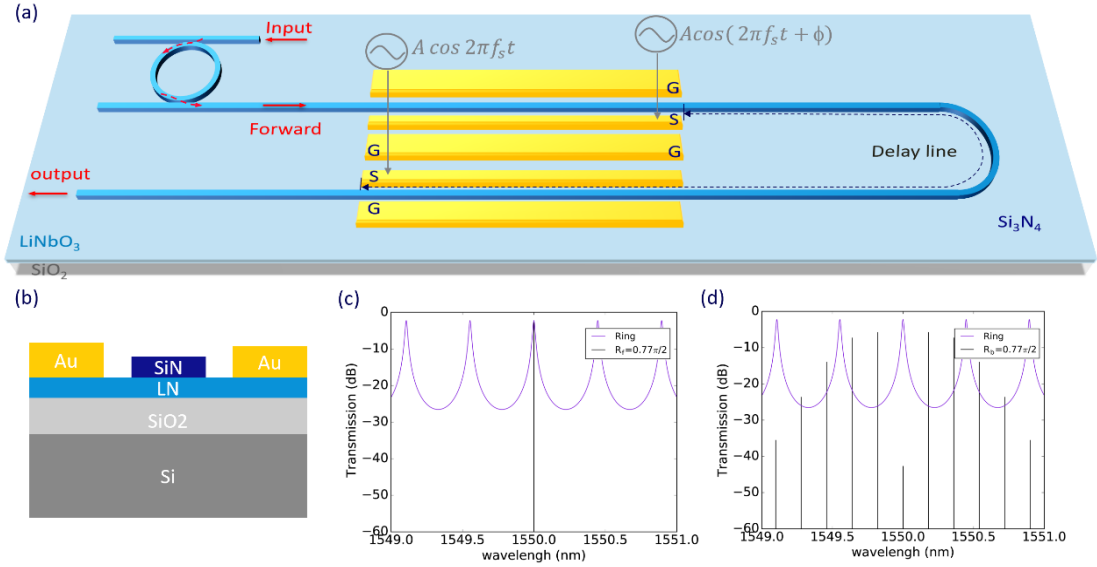

*Fig. 1. (a) Illustration of the investigated spatiotemporal isolator in LNOI. (b) Cross-section of the SiN loaded LNOI waveguide. (c) Purple curve shows the simulated spectral response of the add-drop ring resonator with an FSR of 55GHz. Black line is simulated spectral response for light propagating in the forward direction for a modulation index of $0.77\pi/2$. (d) Simulated spectral response for the reverse direction assuming a modulation index of $0.77\pi/2$.*

## 3. Experimental Results and Discussion

To experimentally demonstrate the designed isolator, we fabricated the device in the silicon nitride (SiN) loaded thin-film lithium nobate on insulator integrated waveguide platform. We ordered the wafers from NanoLN and chose X-cut LNOI, to make use of LN's strongest electro-optic tensor component r33, for Y-propagating modulators. The details of the waveguide and electrode fabrication steps are outlined in our prior work [19]. An optical microscope image of the fabricated device is presented in Fig 2(a), showing the input at the top left corner, which couples to an add-drop racetrack resonator before passing through the two cascaded phase modulators. The add-drop ring resonator has a separation (edge to edge) of 1.05 µm between the bus waveguide and coupler. An optical microscope image of the coupling region is shown in Fig 2(b). For the modulator design, whose segment is presented in Fig. 2(c), the length of the travelling wave modulator is 7060 µm with a 5.8 µm electrode gap and a 0.5 µm thick Au electrode. The width and height of the SiN loading waveguide, shown in the scanning electron microscopy (SEM) image in Fig. 2(d) is 1 µm and 0.3 µm respectively, resulting in single mode operation at a 1550 nm wavelength, similar to our previous demonstrations [19].

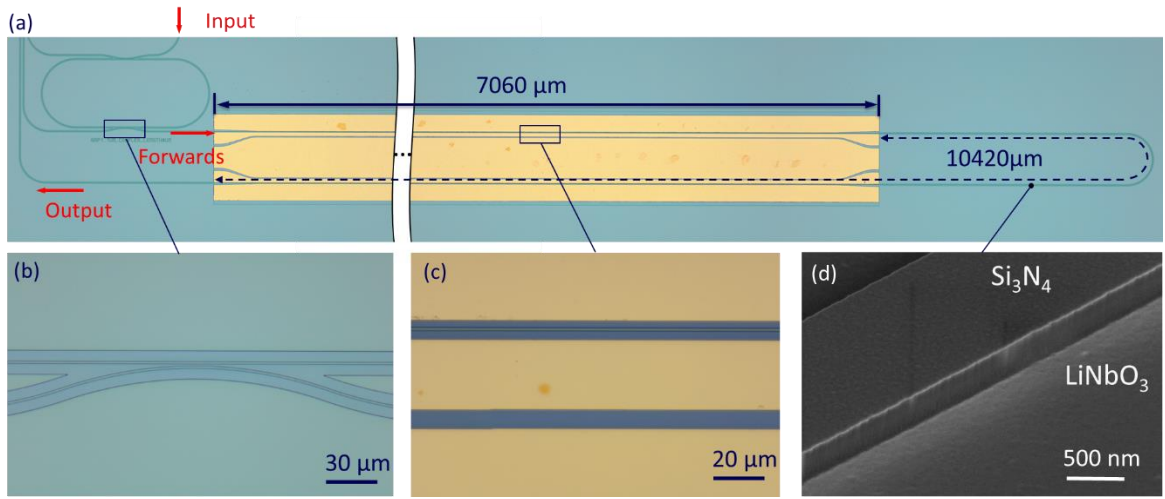

*Fig. 2 (a) Optical microscope image of the fabricated isolator; (b) magnified view of the racetrack resonator coupling region; (c) the traveling wave electrode alignment to the waveguide; and (d) SEM image of the fabricated SiN loaded LNOI waveguide.*

To experimentally characterize the proposed isolator, we first investigated the filtering behavior of the ring cavity using a tunable laser. The drop-port transmission spectra of the ring resonator is shown as a blue line in Fig. 3(a). Afterwards, we characterized the modulator without the ring resonator. For this we coupled 1550 nm laser light such that we operate the device in reverse direction and measured the spectra with a Finisar WaveAnalyzer 1500S High-Resolution Optical Spectrum Analyzer when a 22.5 GHz sinusoidal and cosine signal with the same power from an arbitrary waveform generator (Keysight M9505A) is applied to respective traveling wave electrodes. The microwave signals were injected in the direction of on-chip reverse propagating light, such that the light experiences strong modulation due to being velocity matched. On the opposite side of the traveling wave electrodes microwave signals were terminated using 50 Ω impedance matching resistors. The power of the modulating RF signals was adjusted until the power of the carrier was minimal. In Fig. 3(a), the red curve shows the spectra for the reverse propagating light, when 20.43 dBm microwave power is applied onto both modulators. One can see that, the optical power in the carrier has been suppressed and most of the power has been transferred to the sidebands.

To validate the isolator operation and characterize the effectiveness of the ring resonator in suppressing the undesired optical power in the sideband, we aligned the laser wavelength so that it is on resonance with the racetrack resonator and analyzed reverse-propagating light output at the drop port. The spectrum of the filtered light is shown in Fig. 3(b). Compared to Fig.3(a) one can see that the sidebands are strongly suppressed through spectral filtering.

Next, we measured the spectra when operating the optical isolator in the forward direction and passing the light through the ring resonator (when operating on resonance) with the same RF modulation signals as in the reverse direction. The spectrum is shown in Fig. 3(c). One can see that in this case most of the optical power remains confined in the carrier and only minimal power is transferred to the sidebands and the extra spectral components can be filtered out by another ring resonator or filter. The incomplete suppression of the sidebands is likely caused by the imbalance in the modulation index applied to the two modulators due to the different modulation efficiencies of the phase modulators as a result of fabrication tolerances (e.g., waveguide to electrode alignment). To quantify the isolation strength, we also measured the power difference with an optical power meter when operating the isolator in the forward and reverse directions, resulting in an isolation of 27 dB.

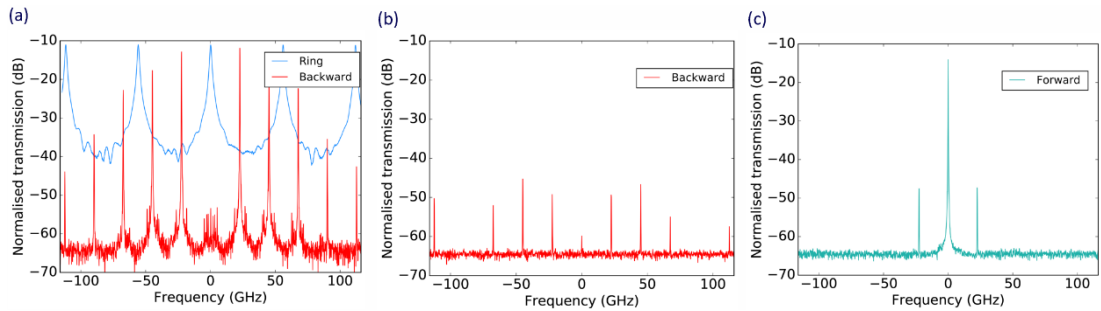

*Fig. 3. (a) the blue curve shows the measured transmission spectrum of the ring resonator. Red curve shows the measured spectrum of the cascaded phase modulators for light travelling in the reverse direction; Measured spectra of the isolator when operating the device in the (b) reverse direction and (c) forward direction, when using the racetrack resonator to suppress sidebands.*

It is important to note that the experimental results for operating the fabricated device in reverse direction showed that after filtering the carrier has less power, compared to the power of the filter-suppressed sidebands. This indicates that the isolation of the device is currently limited by the quality-factor of ring resonator, and it can be improved further by using cascaded ring resonators. One disadvantage of the demonstrated isolator is that the operation wavelength of the light sources needs to align with the resonance wavelength of the ring resonator, making it exceedingly narrowband.

However, additional versatility can be attained by tuning the resonance wavelength of the ring resonator using heaters and locking it to the laser wavelength.

## 4. Conclusion

In this work, we designed and experimentally demonstrated an integrated isolator that is based on spatiotemporal modulation with 27 dB isolation in the LNOI waveguide platform. This was achieved by using LN's excellent electro-optic modulation properties and suppressing the optical power in the sidebands with a ring resonator filter. The isolation ratio of our device was limited by the ring resonator, providing scope for further improvements in the near future, making the demonstrated isolator suitable for future applications such as LNOI circuits with integrated light sources and amplifiers.

## Funding

This work was supported by the Australian Research Council (ARC) grants DP190102773, DP190101576, DP220100488.

## Acknowledgments

The authors acknowledge the facilities and the scientific and technical assistance of the Micro Nano Research Facility (MNRF) and the Australian Microscopy & Microanalysis Research Facility at RMIT University. This work was performed in part at the Melbourne Centre for Nanofabrication (MCN) in the Victorian Node of the Australian National Fabrication Facility (ANFF).

## Competing interests

The authors declare no conflicts of interest.

## Data availability

Data underlying the results presented in this paper are not publicly available at this time but may be obtained from the authors upon reasonable request.